# Coherent surface-to-bulk vibrational coupling in the 2D topologically trivial insulator Bi$_2$Se$_3$ monitored by ultrafast transient absorption spectroscopy

Yuri D. Glinka [1,2,*], Tingchao He [3,*], Xiao Wei Sun[1,4,*]

[1] Guangdong University Key Lab for Advanced Quantum Dot Displays and Lighting, Shenzhen Key Laboratory for Advanced Quantum Dot Displays and Lighting, Department of Electrical and Electronic Engineering, Southern University of Science and Technology, Shenzhen 518055, China
[2] Institute of Physics, National Academy of Sciences of Ukraine, Kyiv 03028, Ukraine
[3] College of Physics and Energy, Shenzhen University, Shenzhen 518060, China
[4] Shenzhen Planck Innovation Technologies Pte Ltd., Longgang, Shenzhen 518112, China

Ultrafast carrier relaxation in the 2D topological insulator (TI) Bi$_2$Se$_3$ [gapped Dirac surface states (SS)] and how it inherits ultrafast relaxation in the 3D TI Bi$_2$Se$_3$ (gapless Dirac SS) remains a challenge for developing new optoelectronic devices based on these materials. Here ultrashort (~100 fs) pumping pulses of ~340 nm wavelength (~3.65 eV photon energy) were applied to study ultrafast electron relaxation in the 2D TI Bi$_2$Se$_3$ films with a thickness of 2 and 5 quintuple layers (~2 and ~5 nm, respectively) using transient absorption (TA) spectroscopy in the ultraviolet-visible spectral region (1.65 - 3.9 eV). The negative and positive contributions of TA spectra were attributed to absorption bleaching that mostly occurs in the bulk states and to the inverse bremsstrahlung type free carrier absorption in the gapped Dirac SS, respectively. Owing to this direct and selective access to the bulk and surface carrier dynamics, we were able to monitor coherent longitudinal optical (LO) phonon oscillations, which were successively launched in the bulk and surface states by the front of the relaxing electron population within the LO-phonon cascade emission. We have also recognized the coherent surface-to-bulk vibrational coupling that appears through the phase-dependent amplitude variations of coherent LO-phonon oscillations. This unique behavior manifests itself predominantly for the topologically trivial insulator phase of the 2D TI Bi$_2$Se$_3$ (2 nm thick film) in the photon energy range (~2.0 – 2.25 eV) where efficient energy exchange between the bulk and surface states occurs. We also found that the coherent surface-to-bulk vibrational coupling significantly weakens with increasing both the Bi$_2$Se$_3$ film thickness and pumping power.

## INTRODUCTION

Nontrivial topology of the band states in some narrow bandgap semiconductors led to the advent of new states of quantum matter with topologically protected helical Dirac surface states (SS) revealing linear dispersion and crossing the material's energy gap.[1,2] Specifically, the robust topological protection of the gapless Dirac SS emerges because strong spin-orbit interaction and time-reversal symmetry act simultaneously, thereby introducing a new class of materials that is usually referred to as topological insulators (TIs).[1,3] Because TIs are expected to be promising materials for potential application in novel optoelectronic and spintronic devices, interest in these materials is constantly growing.

Crystalline rhombohedral bismuth selenide (Bi$_2$Se$_3$) is a great 3D TI candidate for fundamental studies, since it is characterized by a single node structure of the gapless Dirac SS, whereas the bulk remains insulating with the bandgap energy of $E_g$ ~ 0.3 eV.[3] The 3D TI Bi$_2$Se$_3$ is also known as a layered van der Waals structure consisting of the quintuple layers (QLs) of covalently bounded Se-Bi-Se-Bi-Se atomic sheets (QL ~0.954 nm).[1-3] It has also been found that the 3D TI Bi$_2$Se$_3$ reveals the higher energy Dirac SS being energetically distanced from the midgap Dirac SS (SS1) by ~1.5, ~2.7 and ~3.9 eV (SS2, SS3 and SS4, respectively).[4,5] The lowest energy Dirac SS1 are usually partially occupied due to natural $n$-doping,[3] whereas SS2, SS3 and SS4 remain unoccupied and are located between the higher energy unoccupied bulk states. Consequently, the low-energy quasiparticles nearby the SS1 node are known as massless helical Dirac fermions that govern metallic-like surface conductivity in these materials.[6]

On the contrary, the 3D TI Bi$_2$Se$_3$ tends to be fully insulating both in the bulk and on the surface when approaching the 2D limit (below 6 QL film thickness).[7] This behavior results from magnetic coupling between massless Dirac fermions residing on the opposite surfaces of the van der Waals film, thus introducing the energy gap oscillating with the film thickness.[8] The resulting ultrathin Bi$_2$Se$_3$ films with gapped Dirac SS are usually referred to as the 2D TI Bi$_2$Se$_3$.[6,7] Although the gap between the Dirac cones opens progressively with decreasing film thickness, one should distinguish between the gapped topologically nontrivial insulator phase with linear dispersion (5 QL film thickness) and the topologically trivial insulator phase (2 QL film thickness) with nearly parabolic dispersion.[6,7] Consequently, the gapped topologically nontrivial insulator phase presents a quantum system still combining both quantum confinement and topological order with a unique spin texture of the electronic states. Alternatively, it is often taken for granted that the topologically trivial insulator phase is a purely quantum system, for which quantum confinement completely governs the massive fermion dynamics,[8-10] thus allowing the giant Rashba effect to be observed,[11] as in common 2D electron gas (2DEG).[12]

One of the most important aspects in studying the 3D and 2D TI $Bi_2Se_3$ is associated with the coherent surface-to-bulk vibrational coupling. It has recently been suggested that this coupling may strongly enhance the helical Dirac fermion transport in the bulk of the 50 QL thick $Bi_2Se_3$ film due to the in-plane periodic modulation of electronic states, which is induced by a single-cycle THz pulse selectively exciting the polar longitudinal optical (LO) phonon mode [~1.93 THz (~8 meV)].[13] On the other hand, the excitation of coherent LO-phonon oscillations within the ultrafast relaxation of photoexcited carrier has been observed in numerous pump-probe experiments on the 3D and 2D TI $Bi_2Se_3$ using a variety of pump-probe techniques exploiting different wavelength optical pumping from UV to IR and various types of probing, like electron photoemission,[14] light reflection,[15-17] and light absorption.[5,18] Despite the unambiguous assignment of these oscillations to LO phonons,[14-18] the mechanism of their coherent excitation and the role that they play in the vertical surface-bulk-surface carrier transport and in the surface transport of helical Dirac fermions remain unclear and call for further investigations.

Here we repot on the application of chirp-free transient absorption (TA) spectroscopy to study coherent LO-phonon oscillations in ultrathin $Bi_2Se_3$ films with 2 and 5 QL thickness (the topologically trivial insulator phase and the gapped topologically nontrivial insulator phase of the 2D TI $Bi_2Se_3$, respectively). The unintentionally *n*-type doped $Bi_2Se_3$ films were grown using molecular beam epitaxy on the 0.5 mm thick sapphire substrates, being capped afterward by a ~10 nm thick $MgF_2$ protecting layer (see Methods).[19] The ~340 nm wavelength pumping (~3.65 eV photon energy) and a broadband probing from ~1.65 eV to ~3.9 eV were used in the current study. The application of this TA technique naturally complements and extends our previous measurements, in which two-photon IR pumping with photon energy of ~1.7 eV was used,[18] thereby establishing a correlation between the one-photon and two-photon pumping modes. Similar to two-photon pumping, we monitored directly and selectively the transiently excited electron population in the gapped Dirac SS and the bulk states of the 2D TI $Bi_2Se_3$ through the inverse-bremsstrahlung-type free carrier absorption (FCA) and absorption bleaching mechanisms, respectively.

Consequently, the measured pump-probe traces reveal coherent LO-phonon oscillations that associated with the bulk and surface states. We found that the onset of the pump-probe traces, which is usually associated with zero-time, exhibits a temporal shift from the actual zero-time with a decrease in probing photon energy. This behavior is not associated with the temporal chirp of the supercontinuum probing pulse and points to a direct Pauli blocking mechanism of absorption bleaching. The resulting temporal dynamics also unambiguously proves the successive launching of coherent LO-phonon oscillations by the front of the relaxing electron population within the LO-phonon cascade emission. The corresponding phase difference between successively launched LO-phonon oscillations allowed us to observe a unique coherent surface-to-bulk vibrational coupling that appears for the topologically trivial insulator $Bi_2Se_3$ through the phase-dependent amplitude variations. This coupling manifests itself in the energy range (~2.0 - 2.25 eV) where the efficient energy exchange between the bulk and surface states occurs. Because this energy range is well accessible with the commercial ultrafast Ti:Sapphire lasers, the 2D topologically trivial insulator $Bi_2Se_3$ is expected[1] to have a great promise for a wide range of optoelectronic applications. This unique property of the 2D topologically trivial insulator phase is found to be completely suppressed due to the bulk LO-phonon dephasing with increasing both the $Bi_2Se_3$ film thickness and pumping power. We have concluded that it is necessary to distinguish between the bulk and surface states even for the topologically trivial insulator phase of the 2D TI $Bi_2Se_3$, in contrast to the purely quantized energy levels commonly used for the 2DEG model. Finally, we have also proposed that the vertical electron transport in the 2D van der Waals film establishes coherence between LO-phonon modes launched in the individual QLs of the film.

**RESULT AND DISCUSSION**
**TA spectra of 2 and 5 QL thick $Bi_2Se_3$ films.** The chirp-free TA spectra of ultrathin $Bi_2Se_3$ films of 2 and 5 QL (~2 and ~5 nm) thick that present the topologically trivial insulator phase and the gapped topologically nontrivial insulator phase of the 2D TI $Bi_2Se_3$, respectively, which were measured using the ~3.65 eV photon energy pumping of different powers, extend over a wide spectral region from ~1.65 to ~3.9 eV (Figs. 1 and 2). The development of the TA spectra for ~1 ps is a convolution of the true ultrafast carrier relaxation dynamics and the temporal chirp of the supercontinuum probing pulse. This behavior is characteristic of sub-picosecond TA spectroscopy of semiconductors[20] and solutions of organic compounds.[21,22] The corresponding chirp correction result is shown as an example in Figs. 3a-c for the topologically trivial insulator phase (see Methods). Since the high energy edge of the TA spectrum slightly exceeds the pumping photon energy by ~0.2 - 0.3 eV, the excitation of electrons necessarily occurs simultaneously from both the bulk states in the valence band (VB) and from the Dirac SS1 below the Fermi level in the conduction band (CB) (Fig. 3d). The corresponding Fermi energy can be estimated at ~0.2 - 0.3 eV. This behavior is in good agreement with what was previously reported for the 2D and 3D TI $Bi_2Se_3$.[3-5]

First, we note that we did not observe any qualitative difference in ultrafast relaxation trends between the two samples examined. Moreover, we clearly observed the bulk and surface effects for both of those, although the topologically trivial insulator phase is expected to behave like a common 2DEG.[8-10] These observations, combined with our previous reports,[5] clearly show that the surface and bulk states should be distinguished even for film thicknesses well below the 2D limit of the topologically nontrivial insulator phase (6 QL). Consequently, we will continue to characterize our samples in terms of the bulk and surface contributions even for the topologically trivial insulator phase. The wide spectral range of TA spectra points out that the electron energy relaxation occurs through the higher energy bulk quantum states and Dirac SS (Fig. 3d). Here and further below we refer to a band diagram calculated for the 3D TI $Bi_2Se_3$ (6 QL film thickness),[4] despite applying for the 2D TI $Bi_2Se_3$.

In general, there are a total of three contributions to the TA spectra, namely, two negative contributions and one positive contribution (Figs. 1 and 2). Specifically, the first broadband negative contribution gradually develops from its initial position of ~3.9 eV toward the final energy being slightly below ~1.65 eV within ~0.8 ps. This contribution is associated with the CB absorption bleaching that characterizes the relaxation of the transiently excited electron population through the LO-phonon cascade emission in both the surface and bulk states, occurring in the energy range exceeding ~2 eV.[5,18] This behavior appears due to the direct Pauli blocking mechanism as a progressive extension of the relaxing electron population over the entire probing region (Figs. 1a and 2a). We note that this mechanism is not as widespread as the mechanism for changing the complex refractive index of a material due to the integrated filling of the phase space, usually found in pump/probe reflectivity experiments using the same pumping and probing wavelength.[15-17] The latter mechanism, due to the Kramers-Kronig relations, changes the complex refractive index in a wide spectral range and is different from the direct Pauli blocking mechanism, which is spectrally selective and can be recognized mainly in TA spectroscopy. At the lowest pumping power applied, the broadband negative contribution enhances with increasing film thickness (Figs. 1b and 2b), almost approaching that observed for the 3D TI $Bi_2Se_3$ film (~10 QL thickness).[5] In contrast, it is gradually decreased with increasing pumping power. Both dynamics are in good agreement with the progressively dominating influence of Dirac SS in the TI $Bi_2Se_3$ films with decreasing film thickness and increasing pumping power.[5,17,18]

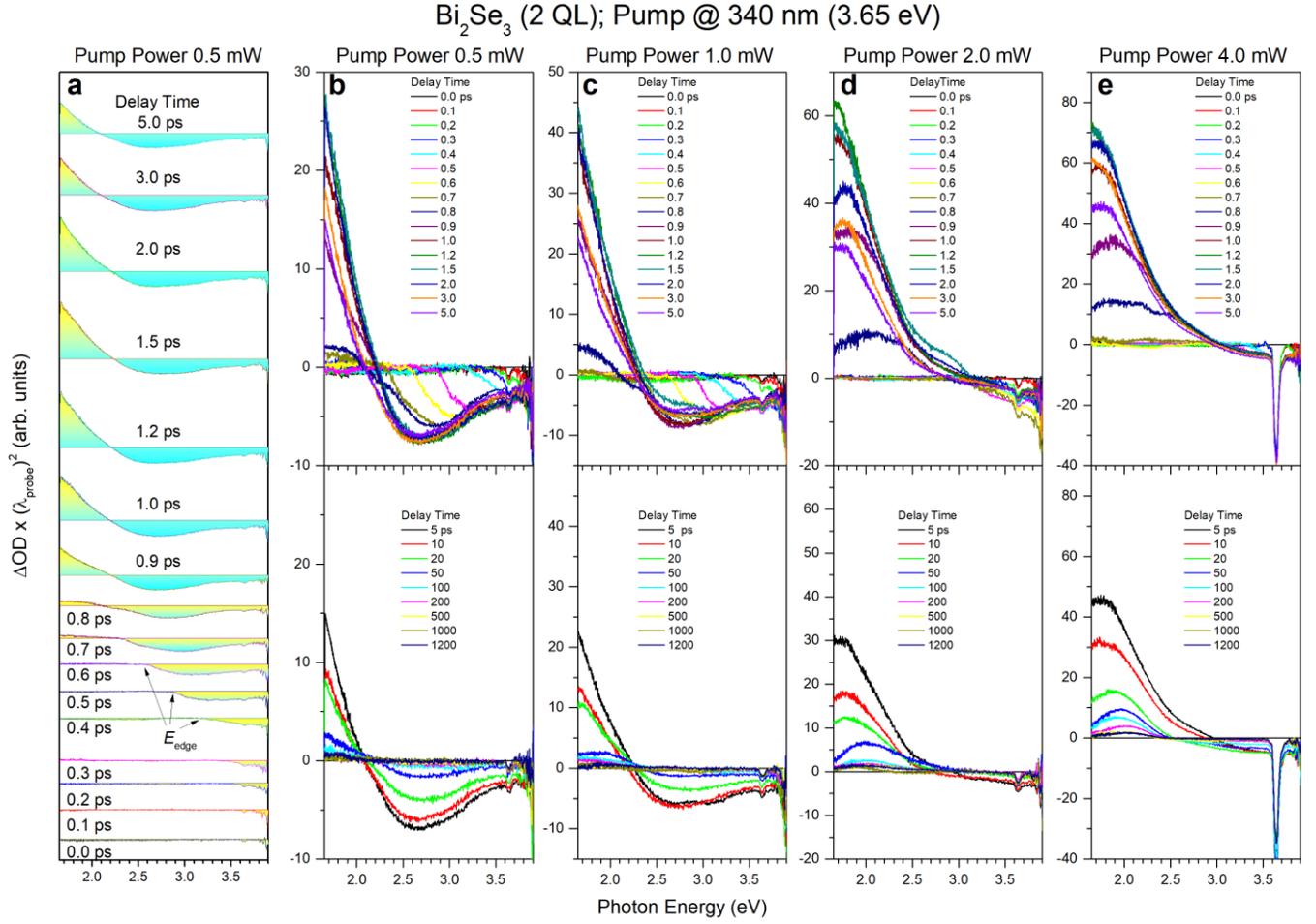

**Fig. 1. TA spectra of the topologically trivial insulator phase of the 2D TI $Bi_2Se_3$.** (a-e) Set of TA spectra of the 2 QL thick $Bi_2Se_3$ film measured at delay times indicated by the corresponding colors using the ~340 nm pumping (~3.65 eV photon energy) of different powers, as indicated for each of the columns. Part (a) and (b) show the same TA spectra for clarity. The zero-intensity lines of TA spectra in part (a) were shifted along the ΔOD axis (optical density change) for better observation. The factor $(\lambda_{probe})^2$ in the ΔOD axis arises due to wavelength-to-energy units' transformation. The low-energy edge of the transiently excited electron population ($E_{edge}$) is indicated in part (a).

The second negative contribution is much narrower and has a maximum nearly matching with the position of the pumping photon energy (~3.65 eV). This contribution is constantly present in all TA spectra, regardless of the delay time, due to the extremely long relaxation dynamics. Consequently, it is associated with the VB absorption bleaching, which mainly characterizes the slow relaxation dynamics of photoexcited holes in the film bulk.[5,18]

Finally, the third contribution is positive and gradually develops from its initial position just below ~1.65 eV to an energy of ~2.25 eV on a time scale from ~0.6 to ~1.5 ps. This contribution is associated with the inverse-bremsstrahlung type FCA and requires the accumulation of transiently excited electron population in the gapped Dirac SS2 (Figs. 1a, 2a, and 3d).[5,18] The resulting dynamic redistribution of electrons from the bulk states toward the Dirac SS2 (vertical electron transport) makes the FCA contribution to be destructively related to the CB absorption bleaching contribution.[18] In other words, one increases as the other decreases. This contribution weakens and narrows with increasing film thickness, while it significantly increases and expands towards a higher energy range with increasing pumping power (Figs. 1 and 2). It is remarkable that at the highest pumping power applied, the FCA contribution for the topologically trivial insulator phase far exceeds the CB absorption bleaching contribution and is maximized at ~2.0 eV (Fig. 1e), thus confirming the surface-related optical response.[18]

To summarize this section, chirp-free TA spectra reveal a complex relaxation of carriers in the 2D TI $Bi_2Se_3$ occurring through the LO-phonon cascade emission and involving both the bulk and surface states. Although the carrier relaxation trends for the topologically trivial insulator phase and the gapped topologically nontrivial insulator phases are qualitatively similar, we have found a remarkable dynamic situation occurring in the topologically trivial insulator phase when the highest pumping power is applied. Specifically, the photoexcited electrons tend to reside predominantly in the surface states, whereas the photoexcited holes preferably occupy the bulk states. In general, this dynamic charge separation develops gradually with decreasing film thickness and becomes dominant for the topologically trivial insulator phase. Note that a similar tendency causes the dynamic gap opening in the Dirac SS of the 3D TI $Bi_2Se_3$ with increasing pumping power.[5]

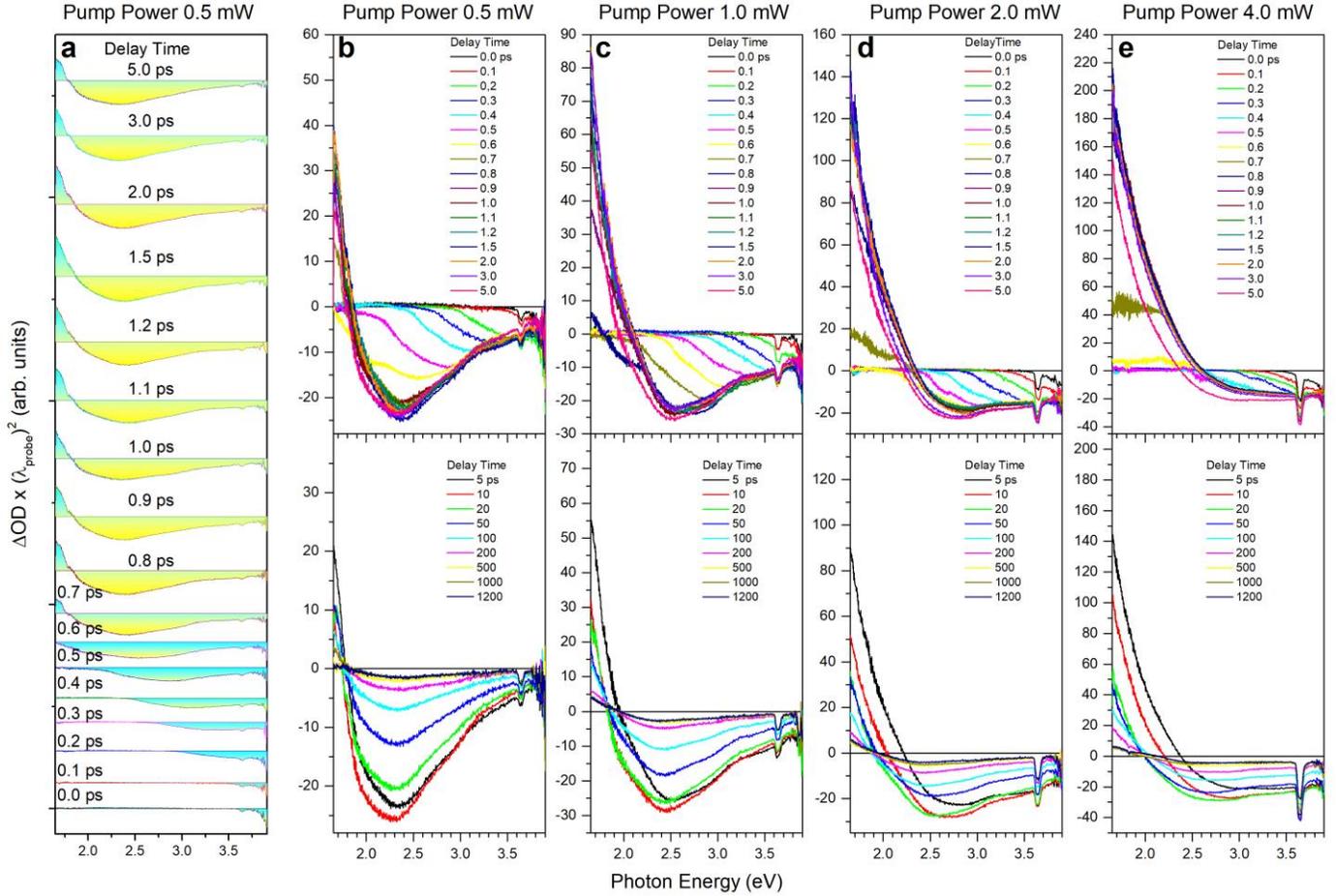

**Fig. 2. TA spectra of the gapped topologically nontrivial insulator phase of the 2D TI $Bi_2Se_3$.** (a-e) Set of TA spectra of the 5 QL thick $Bi_2Se_3$ film measured at delay times indicated by the corresponding colors using the ∼340 nm pumping (∼3.65 eV photon energy) of different powers, as indicated for each of the columns. Part (a) and (b) show the same TA spectra for clarity. The zero-intensity lines of TA spectra in part (a) were shifted along the ΔOD axis (optical density change) for better observation. The factor $(\lambda_{probe})^2$ in the ΔOD axis arises due to wavelength-to-energy units' transformation.

**The electron energy loss rates.** The temporal development of the TA spectra within ∼ 0.8 ps, shown in Figs. 1 and 2, made it possible to estimate the rate of energy loss by electrons using the shift of the low-energy edge of the transiently excited electron population ($E_{edge}$, as defined in Fig. 1a). The resulting temporal dynamics for the topologically trivial insulator phase (Fig. 3e) more clearly demonstrates the two-stage relaxation process compared to the gapped topologically nontrivial insulator phase (Fig. 3f). This two-stage behavior is associated with the relaxation of photoexcited electrons through the cascade emission of LO-phonons, which occurs initially in the higher energy surface states (SS4) and then in the higher energy bulk and other surface states (BS4, BS3, SS3, BS3, BS2), followed by the transient accumulation of electrons again in the surface states (SS2) (Fig. 3d).[5,18] However, this vertical surface-bulk-surface electron transport is strongly dependent on the film thickness and the density of photoexcited electrons. Consequently, the effect of the surface states becomes more pronounced for the topologically trivial insulator phase, thus expanding the corresponding range of delay-times in which this relaxation dominates (Fig. 3e). On the other hand, relaxation occurs predominantly through the bulk states for the gapped topologically nontrivial insulator phase (Fig. 3f). This behavior is very similar to that previously reported for $Bi_2Se_3$ films with a thickness of 4 and 10 QL using the same one-photon pumping.[5] Note also that the two-stage behavior observed for the topologically trivial insulator phase using one-photon UV pumping is similar to that reported earlier for the two-photon IR pumping regime.[18] This coincidence suggests that the dynamics of electron relaxation depends exclusively on the total optical pumping energy. Such a behavior can be achieved when the rates of one-photon and two-photon pumping in Dirac SS

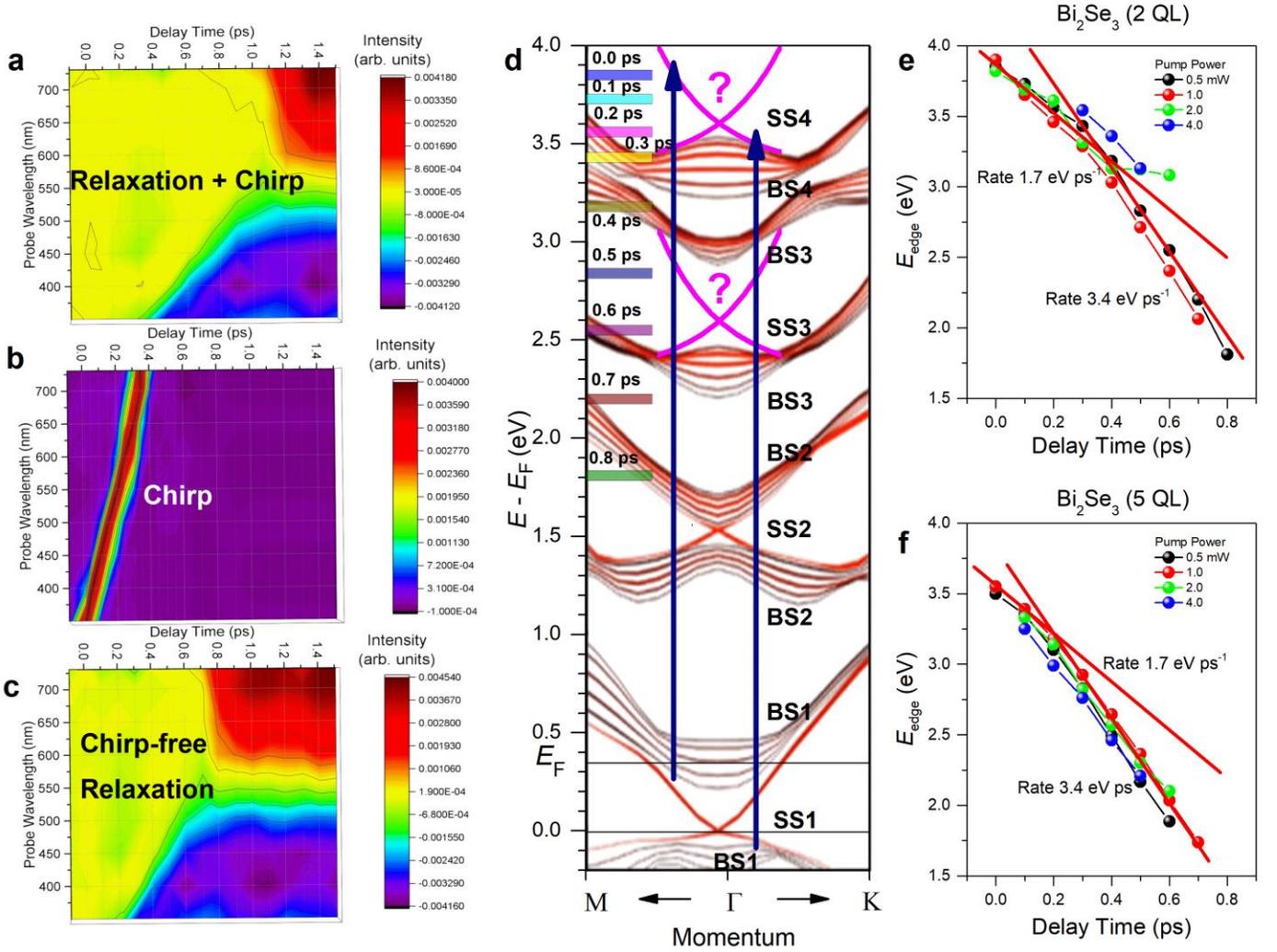

**Fig. 3. Chirp correction, pumping regimes, and electron energy loss rates for the topologically trivial insulator phase and the gapped topologically nontrivial insulator phases of the 2D TI $Bi_2Se_3$.** (a)-(c) Representative pseudocolor TA spectra plots of the topologically trivial insulator phase convoluted with the temporal chirp of the supercontinuum probing pulse, the temporal chirp itself measured with the pumping pulse at ~730 nm (~1.7 eV), and the resulting chirp-free TA spectra, respectively. (d) Band structure of the 6 QL thick 3D TI $Bi_2Se_3$ film calculated in ref. 4 and the ~340 nm (~3.65 eV photon energy) pumping transitions (blue vertical up arrows) originating from the valence band states and from the surface states below the Fermi energy ($E_F$), respectively. The bulk and surface states are marked as BS and SS, respectively. The predicted in ref. 5 higher energy Dirac SS are marked as "?". (e) and (f) Temporal shift of the low-energy edge of transiently excited electron population ($E_{edge}$, as defined in Fig. 1a) for the topologically trivial insulator phase and the gapped topologically nontrivial insulator phases of the 2D TI $Bi_2Se_3$, respectively, measured using the ~340 nm pumping of different powers, as indicated by the corresponding colors. The linear fits (red color straight lines) and the corresponding electron energy loss rates in eV ps$^{-1}$ units are shown. The multicolor bars and the corresponding numbers in ps shown in part (d) present the same data as in part (e) for the 0.5 mW pumping power.

are comparable. We associate this remarkable feature with the giant nonlinearity of 2D Dirac systems.[23]

The corresponding electron relaxation rates are $R_S$ ~1.7 eV ps$^{-1}$ for the surface states and $R_B$ ~3.4 eV ps$^{-1}$ for the bulk states. Furthermore, the carrier relaxation time is in good agreement with the Frohlich relaxation mechanism,[18] significantly exceeding the temporal chirp of the supercontinuum probing pulse (Figs. 3a-c). The twice slower relaxation of electrons in the surface states is due to the much weaker electron-phonon coupling associated with quasi-elastic scattering, compared with inelastic scattering in the bulk states.[18] The quasi-elastic electron-phonon coupling at the surface can be represented as successive electron momentum scattering on the uppermost Se layer of QL, when a small or negligible change in energy occurs. Therefore, this scattering is of a collisional nature, suggesting the existence of a set of surface states converging to the vacuum level. This behavior is also in good agreement with the collisional nature of the inverse-bremsstrahlung type FCA.

**Coherent LO-phonon oscillations and their successive launching.** As discussed in the preceding sections, the dynamics of ultrafast relaxation in the 2D and 3D TI $Bi_2Se_3$ ultrathin films looks similar, nevertheless, demonstrating the predominant effect of the surface states with decreasing film thickness and increasing pumping power. However, the topologically trivial insulator phase of the 2D TI $Bi_2Se_3$ additionally exhibits a unique coupling regime between coherent LO-phonon oscillations associated with the surface and bulk states. We refer to it as the coherent surface-to-bulk vibrational coupling and discuss in the next section. Here, we first consider uncoupled coherent LO-phonon oscillations in the bulk and surface states. Specifically, the emerging picture on ultrafast electron relaxation in the 2D TI Bi$_2$Se$_3$ suggests the successive launching of LO-phonon oscillations by the front of the relaxing electron population within the LO-phonon cascade emission. This behavior clearly manifests itself in the pump-probe traces (Figs. 4 and 5) as a temporal shift of their onsets and the corresponding LO-phonon progressions from the actual zero-time. The onset of pump-probe traces depends hence on which wavelength [or probing photon energy ($E_{probe}$)] this response is monitored at (Fig. 4c). We strongly emphasize the importance of this observation, since the onset of pump-probe traces measured at the same pumping and probing wavelength is usually associated with the cross-correlation time of the corresponding pulses, thus setting the zero-time of the relaxation process. In contrast, if TA spectroscopy is used, the resulting temporal shift ($\Delta t$) with decreasing probing photon energy ($E_{probe}$) images the initial relaxation of photoexcited electrons toward the lower energy states, prior to recombination.

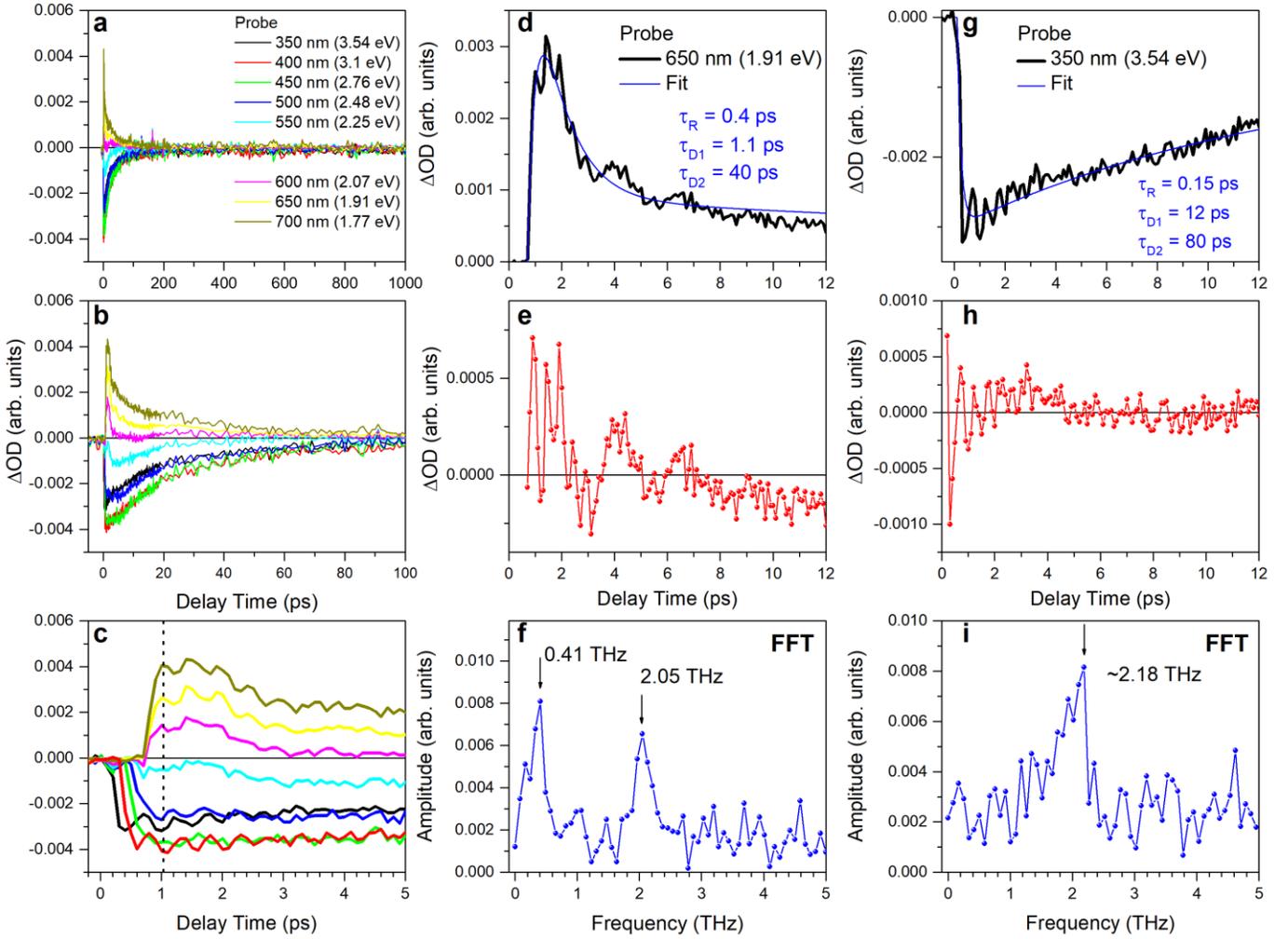

**Fig. 4. The pump–probe traces of the topologically trivial insulator phase of the 2D TI Bi$_2$Se$_3$.** (a), (b), (c) The pump–probe traces for the 2 QL thick Bi$_2$Se$_3$ film, which were measured with the 340 nm (3.65 eV) pumping of 0.5 mW power at the specific probe wavelengths, as indicated by the corresponding colors, and plotted using different timescales. The vertical dotted line in (c) indicates the antiphase character of LO-phonon oscillations for traces measured with $E_{probe}$ = 3.54 and 1.77 eV. (d) and (g) Some of the traces shown in parts (a), (b), (c) and the corresponding fits, as indicated. The corresponding rise-time and decay-time constants are listed for both panels. (e), (h) and (f), (i) The extracted oscillatory parts of the pump–probe traces shown in parts (d) and (g) and the corresponding fast Fourier transformation (FFT), respectively. The center frequencies are indicated in parts (f) and (i).

This relaxation dynamics allows coherent LO-phonon oscillations to appear successively in all pump-probe traces measured within the broadband absorption bleaching and FCA bands in TA spectra. If the resulting coherent LO-phonon oscillations remain collinear, a condition at which all successively launched oscillations occur along the same direction, the corresponding phase difference will control their relative amplitudes. As we found earlier, the rapid dephasing of LO phonon oscillations with a constant of ~1.8 ps, observed for the 3D TI Bi$_2$Se$_3$ films,[17] is accompanied by the vertical electron transport from the bulk states to the Dirac SS2.[18] The latter process usually appears in the pump-probe traces as the initial fast decay stage that reveals a decay-time constant of ~1.2 ps for the thinnest films. The similarity of time constants suggests that coherent LO-phonon oscillations are launched in the presence of a relaxing electron population, which gradually redistributes from the bulk states toward the Dirac SS2 along the normal, i.e., direction perpendicular to the plane of the film (Fig. 3d). This behavior is well consistent with the out-of-plane atomic displacements involved into the $A_{1g}^1$ LO-phonon mode (~9 meV)[24,25] usually appearing in the pump-probe traces.[14-18] Thus, the vertical electron transport apparently establishes coherence between LO-phonon modes that were excited in the individual QLs of the 2D van der Waals film and finally manifested themselves as coherent LO-phonon oscillations in the pump-probe traces. On the contrary, if electrons relax preferably through the bulk states that significantly exceed in energy the Dirac SS2, the initial fast decay stage and coherent LO-phonon oscillations do not appear in the pump-probe traces measured at the corresponding probing photon energies.[18]

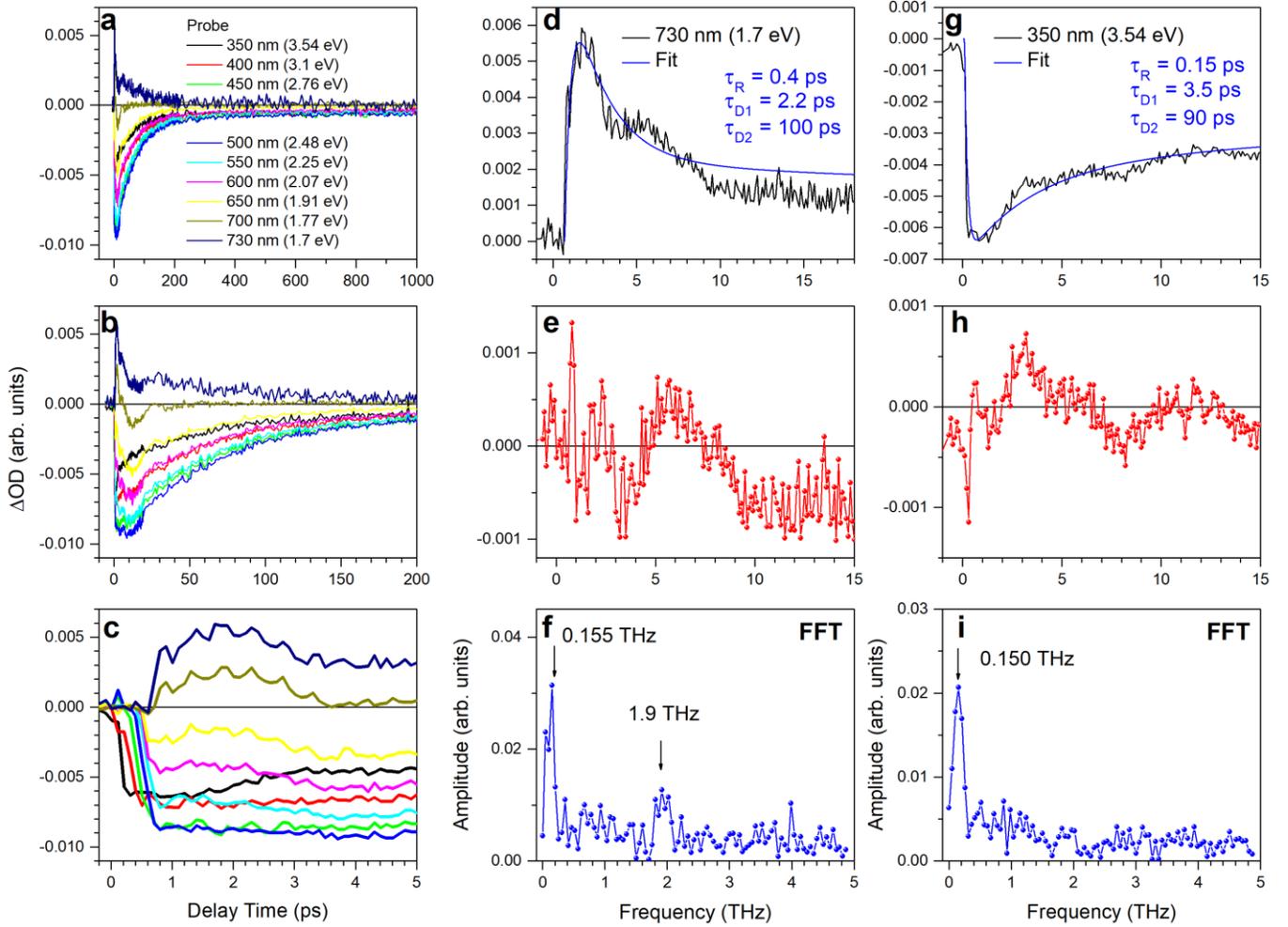

**Fig. 5. The pump–probe traces of the gapped topologically nontrivial insulator phase of the 2D TI $Bi_2Se_3$.** (a), (b), (c) The pump–probe traces for the 5 QL thick $Bi_2Se_3$ film, which were measured with the 340 nm (3.65 eV) pumping of 0.5 mW power at the specific probe wavelengths, as indicated by the corresponding colors and plotted using different timescales. (d) and (g) Some of the traces shown in parts (a), (b), (c) and the corresponding fits, as indicated. The corresponding rise-time and decay-time constants are listed for both panels. (e), (h) and (f), (i) The extracted oscillatory parts of the pump–probe traces shown in parts (d) and (g) and the corresponding fast Fourier transformation (FFT), respectively. The center frequencies are indicated in parts (f) and (i).

To adapt this general picture to the 2D TI $Bi_2Se_3$ films, we first note that the vertical electron transport for these film thicknesses becomes negligible. Consequently, the initial fast decay stage in the pump-probe traces associated with the CB absorption bleaching is significantly weakened (Figs. 4 and 5). At the same time, the amplitude of coherent LO-phonon oscillations for the topologically trivial insulator phase is large enough, whereas the oscillations are significantly suppressed for the gapped topologically nontrivial insulator phase. Since the intrinsic coherence length in QL is limited by its thickness (~1 nm), coherent oscillations of LO-phonon modes are still retained for the thinnest film (~ 2 nm thick). On the contrary, for thicker films (~5 nm), oscillations are suppressed due to the dephasing of LO-phonon modes in the film bulk, being probably caused by their discrete nature in layered van der Waals structures.[26] A further increase in the film thickness (for example, to ~10 nm) restores the oscillation again, since the vertical electron transport becomes strong enough to ensure coherence between LO-phonon modes excited in individual QLs. This kind of nonmonotonic behavior with the film thickness is expected to be strongly dependent on the photoexcited carrier density, thus leading to the optional appearance of coherent LO-phonon oscillations in the pump-probe traces of ultrathin $Bi_2Se_3$ films.[17]

In contrast, the onsets of the pump-probe traces associated with the FCA contribution begin to appear at $\Delta t$ ~0.6 ps and reveal an insignificant temporal shift with $E_{probe}$ (Fig. 4c and 5c). This behavior suggests that the further relaxation of electrons through the gapped Dirac SS2 is less efficient, thus providing their transient accumulation in these states (Fig. 3d). Therefore, it is expected that the resulting phase difference between coherent LO-phonon oscillations launched successively in the Dirac SS2 will be negligible, providing little effect on their amplitudes in pump-probe traces.

The estimated frequencies of coherent LO-phonons nearly match with those known for the bulk and surface states in the 2D and 3D TI Bi$_2$Se$_3$ [$\nu_B$ = ~2.18 THz (~9.02 meV) and $\nu_S$ = ~2.05 THz (~8.48 meV), respectively] (Figs. 4f and 4i).[14,18] The corresponding periods of oscillations are $T_B$ = ~460 fs and $T_S$ = ~490 fs, respectively. The LO-phonon softening effect for the Dirac SS2 suggests that the electrons interact (collide) with the uppermost Se layer of QL. The observed total temporal shift in the onsets of the pump-probe traces measured with $E_{probe}$ = 3.65 and 1.77 eV can be estimated at $\Delta t$ ~1.5 $T_B$ (~3$\pi$). It is noteworthy that for the gapped topologically nontrivial insulator phase (5 nm thick film), the amplitude of the bulk LO-phonon oscillations significantly decreases due to dephasing, whereas the surface LO-phonon frequency slightly lowers (~1.9 THz) (Figs. 5f and 5i). The FCA traces also demonstrate the lower frequency coherent oscillations with frequencies of ~0.41 and ~0.15 THz for the 2 and 5 QL thick 2D TI Bi$_2$Se$_3$, respectively (Fig. 4f, 4i and 5f and 5i). The frequencies of these oscillations strongly depend on the density of photoexcited carriers (Supporting Information), and therefore they can be classified as coherent acoustic Dirac plasmons by analogy with those observed in the 3D TI Bi$_2$Se$_3$.[27]

The amplitude of two collinear LO-phonon coherent oscillations launched successively in the bulk states and measured at a certain $E_{probe}$ can be expressed as,

$$I_B = [sin(2\pi\nu_B t + \varphi_{B1}) + sin(2\pi\nu_B t + \varphi_{B2})]exp\left(\frac{t}{\tau_{LO}}\right)$$
$$= 2A(\varphi)sin\left(2\pi\nu_B t + \frac{\varphi_{B1}+\varphi_{B2}}{2}\right)exp\left(\frac{t}{\tau_{LO}}\right) \quad (1)$$

where $\varphi_{B1} = 2\pi\nu_B \frac{E_{pump}-E_{probe1}}{R_B}$ and $\varphi_{B2} = 2\pi\nu_B \frac{E_{pump}-E_{probe2}}{R_B}$ are the corresponding phases, $A(\varphi) = cos\left(\frac{\varphi_{B2}-\varphi_{B1}}{2}\right) = cos\left[\frac{\pi\nu_B}{R_B}(E_{probe1} - E_{probe2})\right]$ is the phase-dependent amplitude that vanishes when $\frac{\nu_B}{R_B}(E_{probe1} - E_{probe2}) = (2n + 1)$ with $n$ being any integer, and $\tau_{LO}(\gg \Delta t)$ is the dephasing time constant. This analysis suggests that for the observed temporal shift of the onsets of pump-probe traces (~3$\pi$), the amplitude of LO-phonon oscillations is expected to be zero at least two times ($n = 0, 1$). The corresponding delay-times are $T_B$/2 = 230 fs and 3$T_B$/2 = 690 fs.

The measured transient pump-probe traces confirm this phase matching dynamics (Fig. 4c). Specifically, the absorption bleaching traces measured with $E_{probe}$ = 3.54 and 3.1 eV reveal coherent LO-phonon oscillations that are shifted with respect to each other. Consequently, the amplitude of these oscillations drops down to zero for the pump-probe trace measured with $E_{probe}$ = 2.76 eV ($\Delta t$ ~0.3 ps). Subsequently, the amplitude of LO-phonon oscillations recovers again for the trace measured with $E_{probe}$ = 2.48 eV ($\Delta t$ ~0.5 ps), but with a smaller amplitude due to dephasing. The next drop in the amplitude of LO-phonon oscillations is observed for $E_{probe}$ = 2.25 eV ($\Delta t$ ~0.6 ps). Despite minor discrepancies between the experimental and theoretically predicted delay-times, the observed phase matching dynamics seems to be in quite good agreement with Eq. (1).

**Coherent surface-to-bulk vibrational coupling.** In addition to the separate coherent LO-phonon dynamics occurring in the surface and bulk states of the 2D TI Bi$_2$Se$_3$, the coherent surface-to-bulk vibrational coupling can potentially further complicate the relaxation trends in this 2D van der Waals system. The latter assumption raises the fundamental question regarding whether this coupling characterizes the coherent surface-to-bulk vibrational coupling itself instead of the coherent coupling between the corresponding optical responses. Furthermore, the identified mechanism governing the successive launching of coherent LO-phonon oscillations within the LO-phonon cascade emission also raises the question about collinearity between the surface and bulk LO-phonon modes. It is obvious that to observe the coherent surface-to-bulk vibrational coupling, LO-phonon modes in the bulk and at the surface must be excited along the same direction perpendicular to the film plane. Although for the bulk states, the latter statement has been experimentally proven as the LO-phonon-assisted vertical electron transport,[18] the out-of-plane LO-phonon oscillations in the surface states seems to be unique. On the other hand, the latter behavior is well consistent with the quasi-elastic type of electron-phonon coupling in Dirac SS and with the collisional nature of the inverse-bremsstrahlung type FCA, allowing the surface-related electron dynamics to be monitored directly.

It is worth noting here that the in-plane periodic modulation of surface electronic states by a phonon mode with ~8 meV energy selectively excited with a single-cycle THz pulse has recently been suggested to strongly enhance the helical Dirac fermion transport in the close to equilibrium Dirac SS1 of the 3D TI Bi$_2$Se$_3$.[13] However, because the $A_{1g}^1$ LO-phonon mode with ~9 meV energy is known to be associated with the out-of-plane atomic displacements,[24,25] the coherent surface-to-bulk vibrational coupling is rather expected to strongly suppress the incoherent energy dissipative channel in Dirac SS, thus enhancing the surface helical Dirac fermion transport. More specifically, the coherent surface-to-bulk vibrational coupling will significantly enhance the non-dissipative energy exchange between the bulk and surface states when electrons transiently occupy the higher energy surface states (Dirac SS2).[18] This non-dissipative coupling is also well consistent with a higher nonequilibrium mobility found in the Dirac SS2 compared to equilibrium conductivity in the Dirac SS1.[28]

The resulting amplitude of the collinear surface-to-bulk coherently coupled LO-phonon oscillations may be expressed in a similar to Eq. (1) manner,

$$I_{B+S} = [sin(2\pi\nu t + \varphi_B) + sin(2\pi\nu t + \varphi_S)]exp\left(\frac{t}{\tau_{LO}}\right)$$
$$= 2A(\varphi)sin\left(2\pi\nu t + \frac{\varphi_B+\varphi_S}{2}\right)exp\left(\frac{t}{\tau_{LO}}\right) \quad (2)$$

where $\nu = \nu_B \approx \nu_S$ is the average frequency of LO-phonon oscillations, $\varphi_B = 2\pi\nu \frac{E_{pump}-E_{probe}}{R_B}$ and $\varphi_S = 3\pi$ are the phases of coherent LO-phonon oscillations in the bulk and surface states, respectively, and $A(\varphi) = cos\left(\frac{\varphi_B-\varphi_S}{2}\right) = cos\left[\pi\nu\left(\frac{E_{pump}-E_{probe}}{R_B}\right) - \frac{3}{2}\pi\right]$ is the phase-dependent amplitude that vanishes when $\frac{v}{R_B}(E_{pump} - E_{probe}) = 2n + 2.5$ with $n = 0, 1$. The corresponding delay-times are $2.5 T_B/2 = 570$ fs and $4.5 T_B/2 = 1{,}035$ fs.

Figure 4c demonstrates that the first phase matching condition observed with $E_{probe}$ = 2.25 eV ($\Delta t$ ~0.6 ps) overlaps well with that discussed in the preceding section for coherent LO-phonon coupling in the bulk states. However, the actual intensity of the pump-probe trace measured with $E_{probe}$ = 2.25 eV approaches zero, thus indicating a strong overlap of the negative and positive contributions and hence electron populations occupying the surface and bulk states. Consequently, the amplitude decrease of LO-phonon oscillations seems to be mainly caused by the coherent vibrational coupling between LO-phonon modes simultaneously excited in the bulk and surface states. The initial antiphase character of these oscillations is determined by the opposite displacements of uppermost Se atoms of QL with respect to the plane dividing the bulk and surface states.

It is remarkable that the second drop in the amplitude of coherent LO-phonon oscillations due to the coherent surface-to-bulk vibrational coupling should appear at $\Delta t$ ~1 ps, but it does not, despite antiphase oscillations for the pump-probe traces measured with $E_{probe}$ = 3.54 and 1.77 eV (Fig. 4c). First, we note that the latter behavior indicates the absence of optical coupling between the corresponding pump-probe signals. Secondly, this observation also unambiguously proves that despite the general tendency described by Eq. (2), the coherent vibrational coupling between the bulk and surface LO-phonon oscillations occurs exclusively in a certain energy range where these oscillations are launched simultaneously in the presence of an electron population transiently occupying both the surface and bulk states. Such a specific energy range, where the efficient energy exchange between the bulk and surface states occurs, is caused by the overlap of the density-of-states associated with the surface and bulk states, as discussed previously for the 3D TI Bi$_2$Se$_3$ films.[18] This energy range can be estimated for the 2D TI Bi$_2$Se$_3$ as being extended from ~2.0 to ~2.25 eV. On the contrary, if the electron populations of the surface and bulk states are energetically separated, as that occurs at $E_{probe}$ outside the specified region, the system remains coherently uncoupled. This tendency confirms again that the surface and bulk states should be distinguished even for the topologically trivial insulator phase of the 2D TI Bi$_2$Se$_3$, in contrast to the purely quantized energy levels commonly used for the 2DEG model.

This unique surface-to-bulk coupling is completely suppressed for the gapped topologically nontrivial insulator phase (Fig. 5), mainly due to the dephasing of LO phonon modes in the bulk states, as discussed in the preceding section. In contrast, coherent LO-phonon oscillations still arise in the surface states. This behavior confirms the collisional mechanism of the LO-phonon coherent excitation in the surface states, which is independent of the film thickness. Additionally, as the pumping power increases, the coherent dynamics of LO-phonon oscillations become unresolvable for both the surface and bulk states (Supporting Information). We associate this behavior in the surface states with collision-induced dephasing between the initial and final states of electron scattering.

The observed coherent surface-to-bulk vibrational coupling in the topologically trivial insulator phase of the 2D TI Bi$_2$Se$_3$ manifests itself in the energy range that is well accessible with the commercial ultrafast Ti:Sapphire lasers, thus suggesting that this layered van der Waals structure is a promising material for a wide range of optoelectronic applications.

**CONCLUSIONS**

In summary, we have provided evidence that the coherent surface-to-bulk LO-phonon vibrational coupling in the topologically trivial insulator phase of the 2D TI Bi$_2$Se$_3$ can be observed directly using transient absorption spectroscopy. The effect manifests itself as the phase-dependent amplitude variations of the coherent LO-phonon oscillations that were successively launched in the bulk and surface states by the front of the relaxing electron population within the LO-phonon cascade emission. We have also found that the surface and bulk states should be distinguished even for the topologically trivial insulator phase of the 2D TI Bi$_2$Se$_3$, although it is often considered as a conventional 2DEG system. Because the coherent surface-to-bulk vibrational coupling occurs in a certain energy range (2.0 – 2.25 eV) where the efficient energy exchange between the bulk and surface states occurs, the topologically trivial insulator phase of the 2D TI Bi$_2$Se$_3$ seems to be promising for optoelectronic applications exploiting the commercial ultrafast Ti:Sapphire lasers.

**METHODS**

**Samples.** The Bi$_2$Se$_3$ ultrathin films of 2 and 5 nm thick were grown on 0.5 mm Al$_2$O$_3$(0001) substrates by molecular beam epitaxy, with a 10 nm thick MgF$_2$ protecting capping layer, which was grown at room temperature without having to expose the film to the atmosphere. The samples have been found to be epitaxial and of similar structural quality.[19] The nominal number of QL was accurate to approximately 5%. Furthermore, the films reveal the thickness-dependent *n*-type doping with the free carrier density of ~10$^{19}$ cm$^{-3}$.[25]

**Experiments.** TA spectra were measured using the transient absorption spectrometer (Newport), which was equipped with a Spectra-Physics Solstice Ace regenerative amplifier (∼100 fs pulses at 800 nm with 1.0 KHz repetition rate) to generate the supercontinuum probing beam and with the Topas light convertor for the pumping beam. The TA spectrometer was also modified to suppress all the coherent artifacts from the sapphire substrate appearing in the sub-picosecond timescale. Although the chirp effect in the TA spectrometer has been kept constructively to a minimum using parabolic mirrors to collimate and focus the supercontinuum probing beam, and by optimizing the experimental geometry, the data matrix was corrected numerically for remaining chirp by adjusting zero time for each probe wavelength using a coherent signal (presumably degenerate four-wave mixing or two-photon absorption),[20] measured with the ~730 nm (1.7 eV) pumping from a thin (0.3 mm) sapphire plate. To measure TI samples, we used optical pumping at a wavelength of 340 nm (photon energy 3.65 eV). The supercontinuum probing beam being extended over the range of ~1.65 – 3.9 eV was at normal incidence, whereas the pumping beam was at an incident angle of ∼30°. All measurements were performed in air and at room temperature using a cross-linearly-polarized geometry; the pumping and probing beams were polarized out-of-plane (vertical) and in-plane (horizontal) of incidence, respectively.

The spot sizes of the pumping and probing beams were ∼400 μm and ∼150 μm, respectively. The pumping beam average power ranged from ∼0.5 mW to ∼4.0 mW (the corresponding pumping pulse power density ranged from ∼4.0 GW cm$^{-2}$ to ∼32 GW cm$^{-2}$). The broadband probing beam was of the ∼0.4 mW power, which for the same as the pumping beam bandwidth (∼26 meV) provides the probing pulse power density of ∼0.15 GW cm$^{-2}$. Because the latter value is much smaller than that of the pumping pulse, the probing beam effect on carrier excitation is expected to be negligible.

**ACKNOWLEDGEMENTS**
This work was supported by the National Key Research and Development Program of China administrated by the Ministry of Science and Technology of China (Grant No. 2016YFB0401702), the Shenzhen Peacock Team Project (Grant No. KQTD2016030111203005), and the Shenzhen Key Laboratory for Advanced Quantum Dot Displays and Lighting (Grant No. ZDSYS201707281632549). The authors acknowledge J. Li for help with the laser system operation and S. Babakiray for growing the $Bi_2Se_3$ samples by MBE (under supervision of D. Lederman) using the West Virginia University Shared Research Facilities.

**DATA AVAILABILITY**
The data that support the findings of this study are available from the corresponding author upon reasonable request.

**AUTHOR CONTRIBUTIONS**
Y.D.G. modified and tested the transient absorption spectrometer, built the experimental setup, performed optical measurements, and treated the optical experimental data. The optical measurements were performed in the laboratory hosted by T.H. All authors contributed to discussions. Y.D.G. analyzed the data and wrote this paper. X.W.S. guided the research and supervised the project.

**COMPETING INTERESTS**
The Authors declare no Competing Financial or Non-Financial Interests.